\documentclass[11pt, letterpaper]{article}

\usepackage{arxiv}

\usepackage{times}
\usepackage[colorlinks=true, linkcolor=teal, breaklinks=true]{hyperref}
\usepackage{amsthm}
\usepackage{amsmath,amssymb,amsfonts}
\usepackage{helvet}
\usepackage{graphicx}
\usepackage{enumerate}
\usepackage{pgfplots}
\usepackage{mathtools,relsize}
\usepackage{float}
\usepackage{subfigure}
\usepackage{subcaption}
\usepackage{latexsym}
\usepackage{setspace}
\usepackage{xurl}

\usepackage{algorithm,algorithmic}
\usepackage{graphicx}
\usepackage{textcomp}
\usepackage{xcolor}
\usepackage{multirow}
\usepackage{adjustbox}

\DeclareMathOperator*{\argmax}{arg\,max}

\begin{document}
\title{Optimal Energy-Aware Service Management in Future Networks with a Gamified Incentives Mechanism}

\author{
    Konstantinos Varsos \quad Adamantia Stamou \quad George D. Stamoulis \quad Vasillios A. Siris \\
     Department of Informatics, School of Information Sciences and Technology, \\
     Athens University of Economics and Business \and \\
     \texttt{\{kvarsos, stamouad, gstamoul, vsiris\}@aueb.gr} 
}

\date{}

\maketitle
\begin{abstract}
As energy demands surge across ICT infrastructures, service providers must engage users in sustainable practices while maintaining the Quality of Experience (QoE) at acceptable levels. In this paper, we introduce such an approach, leveraging gamified incentives and a model for user’s acceptance on incentives, thus encouraging energy-efficient behaviors such as adaptive bitrate streaming. Each user is characterized by an environmental sensitivity factor and a private incentive threshold, shaping probabilistic responses to energy-saving offers. A serious-game mechanism based on positive behavioral reinforcement and rewards of the users, due to their inclusion in top-K and bottom-M rankings, fosters peer comparison and competition, thus transforming passive acceptance into active engagement. Moreover, within a Stackelberg game formulation, the video streaming service provider—acting as the strategic leader—optimizes both incentive levels and game parameters to achieve network-wide energy and traffic reductions, while adhering to budgetary constraints. This structured approach empowers providers with proactive, application-level control over energy consumption, offering them measurable benefits such as reduced high-bitrate traffic and increased participation in energy-saving behaviors, while also considering user satisfaction. The results of our simulations show that indeed gamification boosts significantly user participation and energy savings provided that the incentive and game parameters are chosen optimally. 
\end{abstract}


\section{Introduction}\label{sec:Introduction}

The rapid growth of 5G and high–definition media consumption—especially 4K and emerging 8K streaming—has intensified energy demand across access, core, and data–center infrastructures \cite{Hossfeld2023AGE}. Bitrate and video quality in streaming directly influence energy consumption and carbon emissions, creating a trade-off between user experience and environmental impact \cite{Kamiya2024JRC}. Actually, based on typical viewership patterns, the vast majority of total end-to-end energy use (i.e., from the data center to the viewing devices) is consumed by end-user devices and home networking equipment \cite{Kamiya2024JRC}. In anticipation of 6G, managing energy consumption must evolve from reactive monitoring to proactive, incentive–driven control, embedding energy–awareness and greenness, without compromising perceived Quality of Experience (QoE). Achieving this balance requires models that accurately capture user behavior, responsiveness to incentives, and the interplay between individual utility and collective environmental welfare \cite{Damiani2013AnAB,Seger2023ReducingTI}. European research efforts and specifically EXIGENCE \cite{EXIGENCESbD} exemplifies this shift by advancing user-centric incentive models and energy-aware service design in digital infrastructures. 

This paper addresses the challenge of rising energy demand by developing a unified framework for energy management and user incentivization. It integrates the greenness–aware QoE model from \cite{Hossfeld2023AGE} with the serious–game–based incentive mechanism proposed in \cite{Papaioannou2016}. The resulting model empowers service providers to dynamically influence user behavior through personalized incentives —whether monetary or behavioral— that promote energy–efficient streaming and computing practices. Each user is characterized by two key parameters: (i) a personalized greenness factor $\gamma_n$, which reflects their environmental sensitivity in QoE evaluation, and (ii) a private incentive threshold $r_{\min,n}$, indicating the minimum reward needed to compensate for the perceived inconvenience of adopting energy-saving actions (e.g., choosing lower bitrate content).

By incorporating the serious–game structure from \cite{Papaioannou2016}, the framework captures social dynamics and competitive behavior among users via top–$K$/bottom–$M$ ranking schemes, fostering gamified engagement with sustainability goals. In this context, users evaluate whether to perform a green action--such as activating energy-saving mode or selecting a lower bitrate--in response to the offered incentive $r_n$ and anticipated social payoff. Their acceptance is probabilistic, modeled through a sigmoid function that reflects diverse behavioral responses and environmental uncertainty.

On the provider side, incentive policies and game parameters $(K, M)$ are optimized within a Stackelberg game setup to achieve system-wide energy reductions while respecting budgetary constraints. This structured approach equips service providers with fine-grained, application-level control over energy consumption. By leveraging personalized incentives and social mechanisms, providers can effectively guide users toward lower bitrate choices, resulting in reduced traffic and energy use. The framework also supports adaptive strategies based on observed user behavior, ensuring both environmental impact and user satisfaction. Quantifiable outcomes include reductions in energy consumption and increased engagement in energy-saving actions.

The framework is validated using a synthetic population generator and extensive numerical experiments that quantify trade–offs between provider cost, user participation, and aggregate traffic savings. We focus on traffic reduction percentage, which corresponds to a relative reduction in the dynamic portion of energy consumption — specifically, the component that scales with bitrate. This excludes the baseline (idle) energy, which remains constant regardless of traffic volume.
The remainder of the paper is organized as follows. Section \ref{sec:related work} reviews related work on game-theoretic, incentive–driven energy management. Section \ref{sec:model} introduces the unified energy–aware user–acceptance model, while, Section \ref{sec:policies} provides the details for the management policies, the serious–game integration and Stackelberg optimization. Section \ref{sec:evaluation} presents numerical results on energy consumption and traffic reduction, followed by sensitivity analyses. Finally, Section \ref{sec:conclusions} concludes with insights on policy design for sustainable 6G services.

\section{Related work}\label{sec:related work}

To begin with, Hossfeld et al. \cite{Hossfeld2023AGE} showed that small reductions in video quality—from “excellent” to “good”—can lead to substantial emission savings, particularly when users are environmentally conscious. Their model introduced the concept of a “green user,” whose willingness to accept lower quality is captured by a greenness factor. Unlike \cite{Hossfeld2023AGE}, which focuses on static QoE valuation, we integrate stochastic acceptance, dynamic incentive adaptation, and Stackelberg-based optimization to support service providers in managing energy-aware streaming at scale.

The proposed gamified incentive framework for energy-aware service management in media streaming builds upon a robust foundation of game-theoretic approaches developed primarily in the context of smart grids and demand-side energy management. Papaioannou et al. \cite{Papaioannou2016} developed a serious-game framework to promote consumer engagement and energy efficiency in smart grids, employing Stackelberg game theory to model user behavior and optimize incentive schemes. We extend this paradigm to the domain of video streaming, aiming to motivate users to accept lower Quality of Service (QoS) in exchange for reduced energy consumption, while maintaining satisfaction levels. In parallel, Krasopoulos et al. \cite{Krasopoulos2022FlexibilityMF} proposed a demand-response model that uses economic incentives and device-level decisions, with user acceptance modeled probabilistically via a sigmoid function. Building on these foundations, our framework combines personalized incentive design with socially-driven gamification, allowing service providers to adapt offers dynamically based on user responsiveness. 

Saad et al. \cite{Saad2012GameTheoreticMF} provided a seminal overview of game-theoretic methods applied to microgrid systems, highlighting the relevance of Stackelberg and non-cooperative games for modeling hierarchical interactions between system operators and consumers. Our work adapts this hierarchical structure to the digital service domain, positioning the service provider as the Stackelberg leader who strategically influences user behavior under energy and budget constraints. Soliman and Leon-Garcia \cite{Soliman2014GameTheoreticDM} introduced user utility functions and storage-aware strategies in their game-theoretic demand-side management model. Our framework parallels this by defining user utility in terms of QoE and energy awareness, while introducing social gamification as a novel behavioral lever.

Byrne et al.~\cite{Byrne2018EnergyMA} optimized grid energy storage systems, emphasizing coordinated control of physical infrastructure. Our framework complements this by reducing upstream demand through user-side behavioral adaptation, indirectly enhancing grid stability and storage efficiency. Han et al.~\cite{Han2019IncentivizingPC} applied cooperative game theory to promote prosumer coalitions, while Tatarenko and Garcia-Moreno~\cite{Tatarenko2014AGT} integrated game and control theory for adaptive, incentive-based demand management. Their focus on feedback aligns with our dynamic tuning of incentives and game parameters for real-time responsiveness. Similarly, Yadati et al.~\cite{Yadati2013IncentiveCM} developed incentive-compatible mechanisms for equitable power cut allocation. Our approach builds on these efforts by embedding equity and transparency into gamified incentive design, ensuring balanced energy-saving outcomes without undermining user satisfaction.

In summary, while prior research has primarily focused on physical energy systems, our contribution lies in translating and extending these game-theoretic principles to the digital service layer. By integrating behavioral incentives, social dynamics, and system-level optimization, our framework enables scalable, user-centric energy efficiency in future 6G networks.

\section{Model}\label{sec:model}

Let $N$ be the set of users, indexed by $n \in \{1, \ldots, N\}$. Each user consumes on-demand digital video and chooses a video resolution between the normal video resolution and the energy-efficient one, which is the green option. We assume that each video has a resolution between a minimum $x_{\min}$ and a maximum $x_{\max}$ bitrates (in kbps).

Let $x_{\ell, n}$, $x_{h, n}$ denote the bitrates of the low and high resolutions for each user, respectively, with $0 \leq x_{\min} \leq x_{\ell, n} < x_{h, n} \leq x_{\max}$. The green action corresponds to selecting $x_{\ell, n}$, and if user $n$ selects it, we call it a green user; it provides a \emph{traffic reduction}  $x_n := x_{h, n} - x_{\ell, n}$. 

From \cite{Hossfeld2023AGE}, we utilize for each user $n$ a greenness factor $\gamma_n > 1$, as an advantage factor, assuming $n$ rates lower video quality with a higher score if it knows that it saves energy. Formally, a green user is satisfied with maximum bitrate $x'_{\max} = x_{\max} / \gamma_n$, and the QoE model from \cite{Hossfeld2023AGE} is depicted as,
\begin{equation}\label{eq:f_gamma}
	f_{\gamma_n}(x) = \frac{4}{\log (x'_{\max}) - \log(x_{\min})}\log (x) + \frac{\log (x'_{\max}) - 5 \log (x_{\min})}{\log (x'_{\max}) - \log (x_{\min})},\end{equation}
which is an increasing function w.r.t to $\gamma_n$. Hence, a larger $\gamma_n$ indicates an increasing intrinsic valuation of energy efficiency. Next, we consider the Mean Opinion Score (MOS) from \cite{Hossfeld2023AGE} as the user's $n$ utility, which we denote with $U_n$, and is normalized in $[0, 1]$. Mathematically, we have 
\begin{equation}\label{eq:utility n user}
	U_n(x) := \frac{1}{5} f_{\gamma_n}(x),
\end{equation}
which is a increasing function w.r.t. to $\gamma_n$. Then the utility loss when user $n$ moves from high to low resolution is defined as $\Delta U_n := U_n(x_{h, n}) - U_n(x_{\ell, n})$. 
Further, we introduce the term $s_n$ to be the savings achieved in the energy bill by user $n$ when watching at bitrate $x_{\ell, n}$. The net benefit loss for user $n$, in monetary units, is then $\text{NB}_{loss, n} = \Delta U_n - s_n$. Following \cite{Minou2015IncentivesAT}, if $r_n \geq \text{NB}_{loss, n}$, then the green option is the optimal decision for the user. Therefore, we define the \emph{minimum acceptable incentive} for user $n$ as  $r_{\min, n} = \max \{\text{NB}_{loss, n}, 0\}$. Ideally, the probability $p_n(r_n)$ for user $n$ to take the green option given offer $r_n$ is $p_n(r_n) = 1$ if $r_n \geq r_{\min, n}$, otherwise we have $p_n(r_n) = 0$. In general, we assume that the provider offers incentives $r_n$ from a set $R$.

To capture the uncertainty of the user about its own utility function at any given time, we model acceptance as a Bernoulli trial; that is, we assume that acceptance is stochastic. In particular, given an offered incentive $r_n$ to user $n$, the probability that user selects the green option now becomes,
\begin{equation}\label{eq:sigmoid}
    p_n(r_n) = \frac{1}{1 + e^{-\delta_n(r_n - r_{\min, n})}},
\end{equation}
where $p_n(\cdot)$ is a smooth, strictly increasing sigmoid function, satisfying $p_n(0) = 0$, $\lim_{r_n \to \infty} p_n(r_n) = 1$, and $p_n(r_{\min, n}) = \frac{1}{2}$. The slope parameter $\delta_n > 0$ controls responsiveness such that increasing $\delta_n$ produces a behavior increasingly closer to the step-function behavior. Now, the utility loss under the stochastic acceptance becomes $p_n(r_n) \cdot \Delta U_n$.

By the construction of the model, the parameter $r_{\min, n}$ is personalized for each user. However, we can consider a simplification in which all videos share the same default resolution and the same alternative resolution, i.e., $x_{h, n} = x_h$ and $x_{\ell, n} = x_\ell$ for all $n \in N$. Under this simplification, we can assume that $r_{\min, n}$ is proportional to the video duration, with the proportionality factor remaining a user-specific parameter. Similarly, for the parameter $\delta_n$, we can also consider that it is the same for all videos to be watched by user $n$.

\paragraph*{Energy consumption and carbon footprint modeling.}
Let $x$ denote the bitrate of a user's streaming session and $T$ the duration of the session (in seconds).
The end-to-end energy consumption of a user $n$ for bitrate $x$ is modeled as
\begin{equation}\label{eq:energy bitrates}
	E_n(x) = \sum\nolimits_{t \in T} P(x,t)\,\Delta t,
\end{equation}
where $P(x,t)$ is the instantaneous end-to-end power consumption of the session, encompassing device, access network, and data center components. Following equation 7 in paper in the \cite{Golard2022EvaluationAP}, we assume that the average power increases sub-linearly with bitrate, and approximate $P(x) = P_0 + \alpha \ x$, where $P_0$ is the static baseline power and $\alpha>0$ captures the scaling of dynamic power with bitrate.
For a session of duration $T$, the total energy consumption is,
\begin{equation}\label{eq:energy bitrates session}
	E(x, t) =  \sum_{t \in T} \sum^{N}_{n = 1}\left(P_0 + \alpha x\right)\Delta t.
\end{equation}
The corresponding greenhouse gas emissions are given by $CO_2(x, t) = \eta \, E(x, t)$, where $\eta = 0.388$ (in gCO$_2$/kWh) \cite{EIA} is the carbon intensity of the energy mix. For simplicity, in both formulas we assume that $T=1$ and $\Delta t = 1$.

For each user $n$, the energy reduction achieved by switching from a high bitrate $x_{h, n}$ to bitrate $x$ is $\Delta E_n(x) := E(x_{h, n}) - E(x)$. Since $E_n$ in equation \eqref{eq:energy bitrates} is a sum of linear functions, we can invert it, so there exists a function $g_n$ s.t. $x = g_n(\Delta E_n)$. Notably, energy reduction leads to lower resolution video streaming, which, in turn, reduces the utility of the users. To capture this, for each user $n$ we map the $\Delta E_n$ to $\Delta U_n$, according to the function $\Delta U_n(\Delta E_n) = U_n(x_{h,n}) - U_n( g_n (\Delta E_n))$. 

\paragraph*{Serious-game extension.} Inspired by \cite{Krasopoulos2022FlexibilityMF}, the provider can consider the users as strategic agents engaged in a serious game designed to promote voluntary energy reduction. For that, the provider publicly announces a ranking of the users based on the energy they consume, that is $E_n(x)$, when they use a video streaming service with bitrate $x$. For $n,m \in N$, we say that user $n$ is higher in the serious-game ranking than user $m$ if $E_n(x_n) \leq E_m(x_m)$. Moreover, the provider sets the number of top and bottom users in the serious-game ranking, denoted as $K$ and $M$, respectively. We call top-$K$ (bottom-$M$) the subset of $N$ with the $K$ top ($M$ bottom) users. 

We focus on a serious game that concerns the energy-consumption reduction on behalf of the users at a specific time slot. From equation \eqref{eq:energy bitrates}, the $E(x_n)$ is the  energy consumption of user $n$ associated with bitrates $x_n$. Clearly, equation \eqref{eq:energy bitrates session} provides the total absolute energy consumption, $E(x_{\max})$ is the maximum energy consumption, and $E(x_{\min})$ is the minimum energy consumption.

For a given population of size $N$, the energy reductions $\Delta E_{n}$ for $n \in \text{top-}K$, and $\Delta E_{n}$ for $n \in \text{bottom-}M$, correspond to the top-$K$ and bottom-$M$ quantiles, respectively. Based on \cite{Krasopoulos2022FlexibilityMF}, for each user $n$ we define a reward coefficient $h_n(\cdot)$ as
\[
h_n(x; K, M) =
\begin{cases}
+H, & \text{if } n \in \text{ top-}K,\\
-H, & \text{if } n \in \text{ bottom-}M,\\
0, & \text{otherwise,}
\end{cases}
\]
where $ H \geq 0$. It is important to note that $H$, $K$ and $M$ are naturally correlated: a user in more restrictive top-$K$ list experiences a larger effective reward $H$, since being in a smaller upper group carries more significance. Similarly, for each $K$, there exists an $M$ for which the negative reward to a user due to its inclusion in the penalized bottom-$M$ equals (in absolute terms), the positive reward $H$ associated with inclusion in the top-$K$ list. In this work, however, we consider $H$, $K$, and $M$ to be disentangled, allowing us to independently explore the effects of $H$, $K$, and $M$. Further investigation of the relation among these three parameters falls off the scope of the present paper. However, in practice, a provider employing our approach should initially perform such a study, and then choose optimally parameter $K$ and subsequently $H$ and $M$.

The strategy for each user $n$ is to adjust its video resolution from $x_{h,n}$ to $x_n$, thus reducing its energy consumption by $\Delta E_n(x_n) \geq 0$. Its performance is affected by both $U_n(x)$ and the reward coefficient $h_n$. Therefore, the total expected utility of the user $n$ is $\widehat{U}_n(x) = U_n(x) + h_n(x)$. Each user $n$ chooses a bitrate $x_n$ to maximize the expected $\widehat{U}_n(x_n)$, which depends on both users' decisions and on the values of the parameters $K$, $M$ set by the provider. 

If a user $n$ accepts the greener option and/or participates in the serious game, it selects a video resolution $x^*_n$ that corresponds to an energy-reduction level $\Delta E_n (x^*_n) \in [\Delta E_n(x_{\max}), \Delta E_n(x_{\min})]$, maximizing $\widehat{U}_n(x_n)$. 

In this setting, the provider takes two actions: first, it selects the values of $K$, $M$, and second, it offers incentives $r_n$ to the users. Now, the provider-induced ``effective incentive'' is $r_n + h_n$ and the acceptance probability becomes
\begin{equation}\label{eq:serious game r_n incentive}
    p_n(r_n;K,M) = \frac{1}{1 + \exp\!\bigl(-\delta_n\bigl(r_n + h_n - r_{\min,n}\bigr)\bigr)}.
\end{equation}

\section{Policies}\label{sec:policies}

In the previous section, we discussed the basic model and the utilities of the users. Now, we turn our attention to the provider. Initially, in our model, the provider offers the incentives $r_n 
\in R$ and then receives the expected total traffic reduction as follows
\begin{align}
\label{eq:expected-flex}
& \mathbb{E}[\mathrm{Traffic \ Reduction}] := \sum\nolimits_{n=1}^N p_n(r_n) \cdot (x_{h,n} - x_n).
\end{align}

Further, assume that the provider has a specific bounded budget $B$, measured in abstract monetary units (MU); therefore, the incentives it can offer must satisfy the condition $\sum_{n=1}^N p_n(r_n) r_n \leq B$,  and the provider aims to maximize the total traffic reduction, which is formalized as follows,
\begin{equation}\label{eq:plo2}
	\begin{array}{cll}
		\max\limits_{r_n \in R} & \sum_{n=1}^N p_n(r_n) \cdot (x_{h,n} - x_n)  \\ [.8em]
		\text{s.t.} & \sum_{n=1}^N p_n(r_n) \cdot r_n \leq B \\ [.8em]
		& r_n \geq 0, \quad \forall n \in \{1, \ldots, N\}.
	\end{array} \tag{P1}
\end{equation}
Now, applying equation \eqref{eq:energy bitrates} we can compute the maximum energy reduction using the formula $\sum\nolimits_{n \in N} p_n(r_n)\cdot \Delta E_n(x_n)$. Further, utilizing the formula $CO_2(x, t) = \eta \, E(x, t)$, as stated above, we can compute the maximum $\text{CO}_2$ reduction using the formula $\sum_n p_n(r_n)\cdot \Delta (CO_2)_n(x_n)$.

In case the provider falls short to cover the total offered incentives due to budget constraints, it utilizes the Algorithm \ref{alg:budget_allocation}. There, the provider promotes the more energy efficient users, lines \ref{al:5}-\ref{al:8}, and split any remaining budget to the rest, line \ref{al:9}.

\begin{algorithm}[H]
\caption{Budget-Constrained Incentive Allocation}
\label{alg:budget_allocation}
\begin{algorithmic}[1]
\REQUIRE Number of users $N$, total budget $B$,  offered incentives $r_n$s, acceptance probabilities $p_n$.
\IF{$\sum_{n = 1}^N p_n(r_n) \cdot r_n \le B$}
    \STATE $\widehat{r}_n \gets r_n$
\ELSE
    \STATE Compute efficiency: $\frac{\Delta E_n}{r_n}$ and sort users by decreasing efficiency in list $L$.
    \FOR{$i \in L$} \label{al:5}
        \STATE Add $p_i(r_i) \cdot r_i$ up to $B$ and store the $i$s.
        \STATE $\widehat{r}_i \gets r_i$
    \ENDFOR \label{al:8}
    \STATE If there is a budget remainder split it equivalently to the remaining users, $\widehat{r}_j$.  \label{al:9}
\ENDIF
\RETURN New offered incentives $\widehat{r}_n$s.
\end{algorithmic}
\end{algorithm}

\paragraph*{Serious-game extension.} 
We model the serious game between the service provider and the users as a Stackelberg game. In the first stage, the provider of the serious game (the leader) chooses the game parameters--specifically the top-$K$ positions, the bottom-$M$ positions, and the offered incentives $r_n \in R$. It aims to convey to the users, so as to minimize the energy consumption of video resolution while steering user behaviour. In the second stage, each user (a follower) reacts to the announced game rules and any offered incentives by selecting the bitrate $x$ that maximizes its own expected utility under the rules of the serious game, given by $\widehat{U}_n(x)$.

Concretely, for given game parameters $(K,M)$, the total energy demand at the time of interest is $E(x; K, M) = \sum_{n = 1}^N  E(x_n; K, M)$. Under our unified model, the user acceptance of an incentive offer $r_n$ is stochastic and depends on the effective incentive $r_n + h_n$. The provider aims to maximize the energy reduction under an incentive policy, which is 
\begin{equation}\label{eq:plo5}
    \begin{array}{cl}
		\max\limits_{K,M,r_n} & \mathcal{E}(x_n,r_n;\,K,M) \\
        \text{s.t.} & \sum_{n=1}^N p_n(r_n) \cdot r_n \leq B \\ [0.8em]
		& r_n \geq 0, \quad \forall n \in \{1, \ldots, N\},
    \end{array} \tag{P2}
\end{equation}
where 
\[
\mathcal{E}(x_n, r_n;\,K,M) := \sum\nolimits^{N}_{n = 1} p_n(r_n; \, K,M) \cdot \Delta E_n(x_n).
\]

The solution concept in this two-stage interaction is the Stackelberg equilibrium, which consists of the provider's strategy $(K^\ast,M^\ast,r^\ast)$ and users' responses $x^{\ast}$, following the rules,
\begin{enumerate}
  \item Given the provider’s strategy $(r_n^\ast,K^\ast, M^\ast)$, each user $n$ maximizes its own utility $\widehat{U}_n(x)$, by picking a bitrate that is best response to $(r_n^\ast,K^\ast, M^\ast)$. We denote this choice as $x^{\ast}$. Therefore, it provides the optimal $\Delta E_n(x^{\ast})$, which we denote it as $\Delta E_n^\ast$.
  \item Given the users' best responses $x^{\ast}$, the provider's choice $(r_n^\ast,K^\ast, M^\ast)$ maximizing the energy reduction, that is problem \eqref{eq:plo5}.
\end{enumerate}
Formally, letting $BR_n(r_n,K, M) = x_n$ denote the best-response mapping of user $n$, and $\mathcal{E}(x_n, r_n;\,K,M)$ the energy reduction evaluated at the users’ best responses, a Stackelberg equilibrium satisfies
\begin{equation}\label{eq:Stack-eq}
\begin{split}
    x_n^\ast = BR_n(r_n^\ast,&K^\ast,M^\ast)\quad\forall n \in \{1, \ldots N\},\\
    (r_n^\ast,K^\ast, M^\ast) & \in\argmax_{r_n,K,M} \; \mathcal{E}(x_n, r_n;\,K,M).
\end{split}
\end{equation}
In case the provider faces budget limitations, it can allocate the offered incentives according to the Algorithm \ref{alg:budget_allocation}.

\section{Evaluation}\label{sec:evaluation}

\subsection{Synthetic data generator}\label{subse:synthetic data generator}
To evaluate the gamified incentive model, we designed a synthetic data generator that produces a population of $N = 1000$ users. Each user's attributes mirror real-world variations in video-streaming behavior, environmental awareness, and incentive responsiveness. At first, \emph{high/low bitrates} $(x_{h,i},x_{l,i})$ are drawn from discrete sets $\{2000,3000,4000,5000\}\;\mathrm{kbps}$ and $\{300,600,1200,1500\}\;\mathrm{kbps}$, respectively. Offered incentives are sampled 
from the normal distribution $\mathcal{N}(\mu, \sigma^2)$. So we examine the case where the provider has little information about the population $N$, that is, only the mean $\mu$ and the variance $\sigma^2$. Given only this limited information about users' preferences, the $\mathcal{N}(\mu, \sigma^2)$ ensures the least-biased incentive spread consistent with the provider’s budget constraint. Further, we vary the parameters $H$, $B$, $\delta_n$ as specified in each experiment, 
and we take $\delta_n \sim \mathcal{U}(1, 10)$, $\gamma_n \sim \mathcal{U}(1, 5)$, $P_0 = 10$, $\alpha = 0.2$, $c_{\text{admin}} = 0.04$ globally, and $r_n \geq 0$. 
Finally, we utilize absolute energy quantities in W/h and report the relative traffic reduction (\%) based on the total baseline traffic, that is $\sum_{n \in N} x_{h, n}$, and budget $B$ in Monetary Units (MU). Moreover, before proceeding to the experimental results, we perform a preliminary step. We sample $1000$ $x_{h,n}$ and $x_{l,n}$ video resolutions bitrates' for each of the $N$ users. Then the average maximum traffic becomes $\sum_{n \in N} x_{h, n} \approx 3.5 \times 10^3$~kbps, while for the average minimum traffic consumption we have $\sum_{i \in [n]} x_{l, n} \approx 0.86 \times 10^3$~kbps. Therefore, given the data for the high/low bitrates, we can achieve at most $75.6\%$ traffic reduction. Also, we exploit the relationship between the traffic reduction and the energy reduction as presented in Section \ref{sec:model} and the evaluations are performed in terms of the total reduction.

\begin{figure}[t]
    \centering
    \includegraphics[scale=0.45]{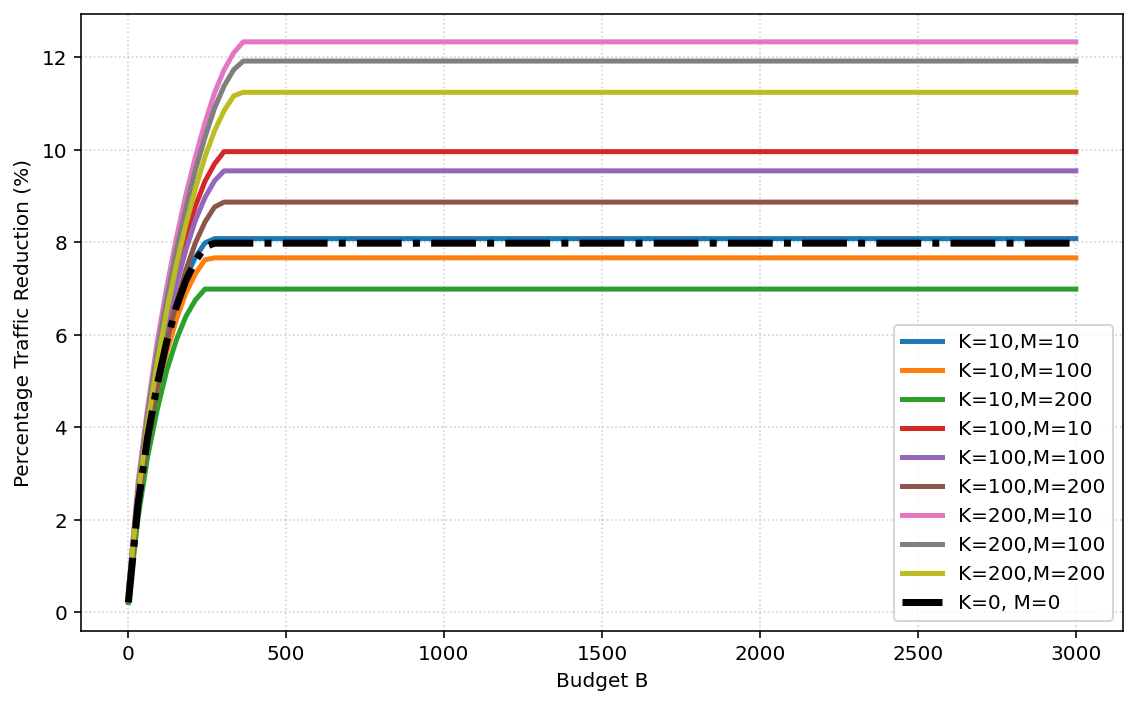}
    \caption{Percentage of traffic reduction with $r_n \sim \mathcal{N}(1,0.25)$ and $H = 1000$.}
    \label{fig:Percentage Traffic Reduction small rn}
\end{figure}

\begin{figure}[t]
    \centering
    \includegraphics[scale=0.45]{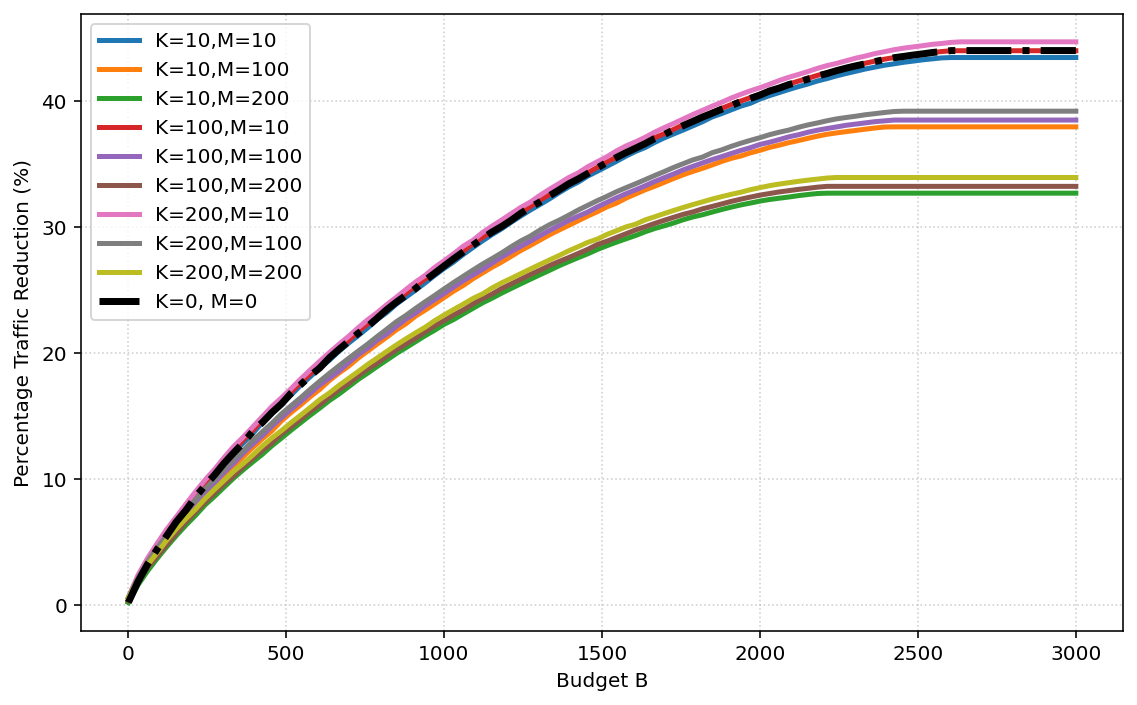}
    \caption{Percentage of traffic reduction with $r_n \sim \mathcal{N}(3, 4)$ and $H = 1000$.}
    \label{fig:Percentage Traffic Reduction medium rn H 1000}
\end{figure}

\begin{figure}[t]
    \centering
    \includegraphics[scale=0.45]{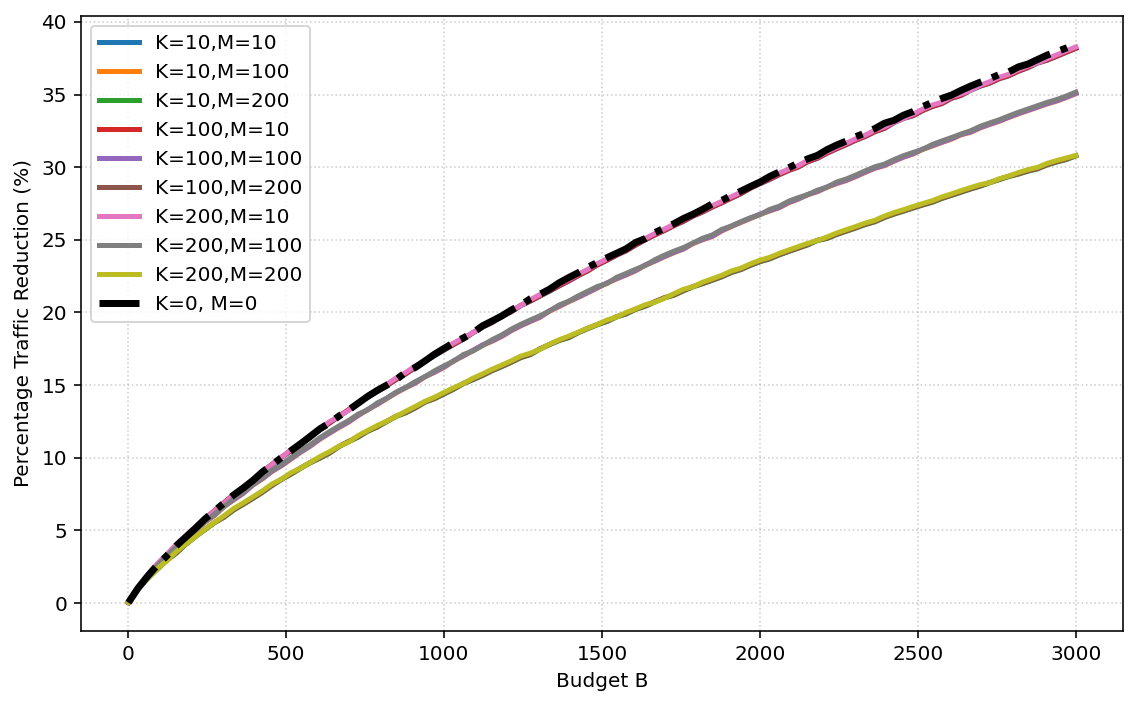}
    \caption{Percentage of traffic reduction for $r_n \sim \mathcal{N}(10, 16)$ and $H = 1000$.}
    \label{fig:Percentage Traffic Reduction large rn H 1000}
\end{figure}

\paragraph*{Effect of serious games parameters $K, M, H$.} In this experiment, the provider has a budget $B \in [1, 3000]$ to offer incentives, and a social reward $H = 1000$ is provided to the top-$K$ and bottom-$M$ lists, with $K, M \in \{0, 10, 100, 200\}$. For the synthetic data described above, the total incentive threshold to be covered is $\sum_{n \in N} r_{\min, n} = 2268$ MU. In Figure \ref{fig:Percentage Traffic Reduction small rn}, 
the black dashed curve ($K=M=0$) shows that without gamification, 
only a $6.5\%$ of traffic reduction, meaning the total offered incentives fall short by 1327 MU. Introducing the top-$K$ and bottom-$M$ lists with the social reward $H = 1000$ increases the traffic reduction to $12\%$ when $K = 200$ and $M =10$. 
When mild incentives are offered, that is $r_n \sim \mathcal{N}(3,2)$, covering only $37\%$ of $r_{\min, n}$s (total shortfall 532 MU), the differences are small and traffic reduction can reach $\approx 40\%$, even without any gamification, see Figure \ref{fig:Percentage Traffic Reduction medium rn H 1000}. The social reward $H$ then further increases the reduction to $\approx 43\%$, or $\approx 58\%$ of the achievable maximum. With higher incentives, that is $r_n \sim \mathcal{N}(10, 4)$, only $3\%$ of $r_{\min, n}$s are not covered, or there is $42$ shortfall. Therefore, the incentive component of our model drives the reduction in bitrates traffic. We observe that even without gamification, see the black dashed curve in Figure \ref{fig:Percentage Traffic Reduction large rn H 1000}, we achieve the best traffic reduction, thereby energy savings, that is $\approx 37\%$.

\begin{figure}[t]
    \centering
    \includegraphics[scale=0.45]{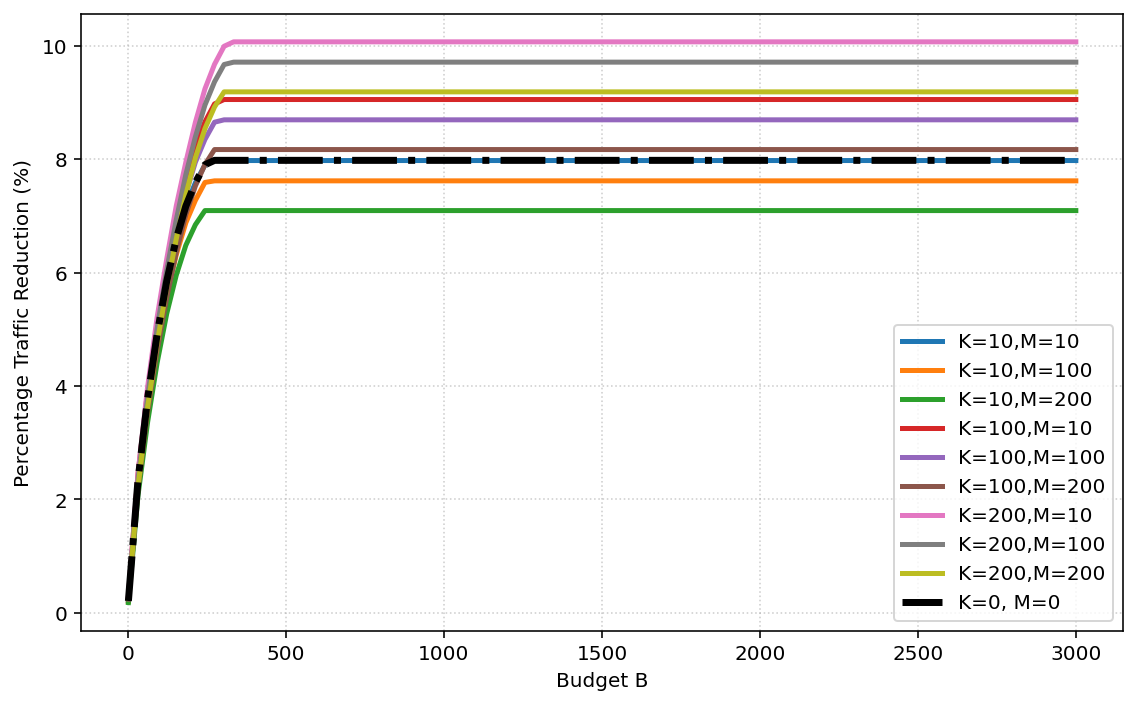}
    \caption{Percentage of traffic reduction for $r_n \sim \mathcal{N}(1, 0.25)$ and $H = 1$.}
    \label{fig:Percentage Energy Reduction small rn H 1 2}
\end{figure}

Figures \ref{fig:Percentage Traffic Reduction small rn}-\ref{fig:Percentage Traffic Reduction large rn H 1000} provide further insights. In particular, across all cases, we observe a similar pattern: settings where $K > M$ perform better, whereas $M > K$ leads to worse outcomes, even compared to no gamification, e.g., see the green curve ($K = 10$, $M = 200$) in all cases. Additionally, the social reward $H$ and the $K$, $M$ values cluster improvements into distinct groups. In Figure \ref{fig:Percentage Traffic Reduction small rn}, the best results occur when $K = 200$, the largest top-$K$ population, because the high $H$ values alter the decisions for all $n \in K$. In contrast, users in the bottom-$M$ list are unlikely to adopt greener behavior, so penalties have minimal effect. In Figures \ref{fig:Percentage Traffic Reduction medium rn H 1000} and \ref{fig:Percentage Traffic Reduction large rn H 1000}, gamification again clusters improvements, but now the critical factor is $M$. Top performance occurs for $M = 10$, this is because the probability to penalize a user that can be incentivized, and so shifted to an un-incentivized user, increases. Larger $H$ values spread the clusters, enhancing settings with large $K$, and compressing those with $K$ small, see Figure \ref{fig:Percentage Energy Reduction small rn H 1 2}. High $H$ accelerates the effect—acting as a multiplier that steepens users’ effective acceptance slope around $r_{\min, n}$. Since the social reward drives the interaction, similar behavior is observed for the other two settings, but we do not present them to avoid repetition. In a punchline, high $H$ allows a lower monetary budget $B$, whereas low $H$ forces the leader to over-provision $B$ to compensate for weak peer effects.

\paragraph*{Effect of budget $B$.} Here, we evaluate the same setting as in the previous experiment. Since the $r_n$s values are small, increasing the budget quickly covers them, causing all the curves in Figure \ref{fig:Percentage Traffic Reduction small rn} to flatten beyond a certain budget. Beyond this point, any remaining budget does not improve outcomes, so the effect of the budget is saturated. For instance, for a total incentive threshold $2268$ MU, the provider can reach $\mathbb{E}[r_n] \approx 1022$ MU; any budget above this yields no further improvement.

Similarly, in Figures \ref{fig:Percentage Traffic Reduction small rn} and \ref{fig:Percentage Traffic Reduction medium rn H 1000}, the budget eventually covers the full $r_n$ range, producing constant behavior, while for higher thresholds (e.g., $\mathbb{E}[r_n] \approx 2900$~MU), the same saturation occurs.

\begin{figure}[t]
    \centering
    \[
    \begin{array}{cc}
    \includegraphics[scale=0.35]{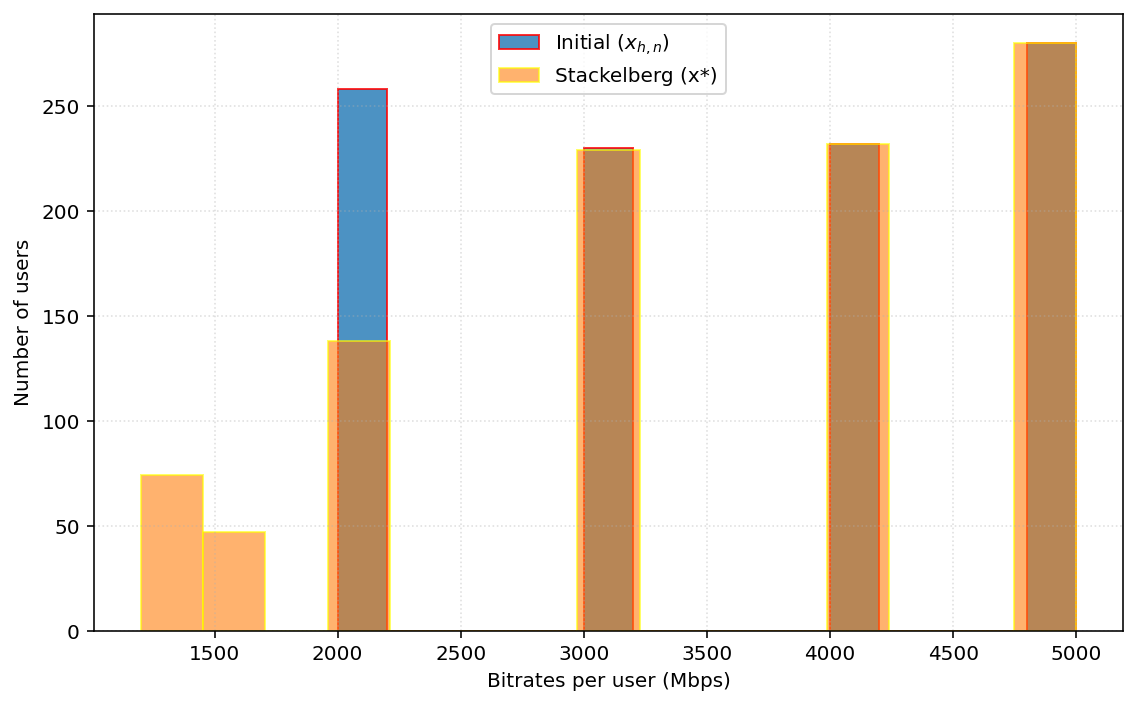} & \includegraphics[scale=0.35]{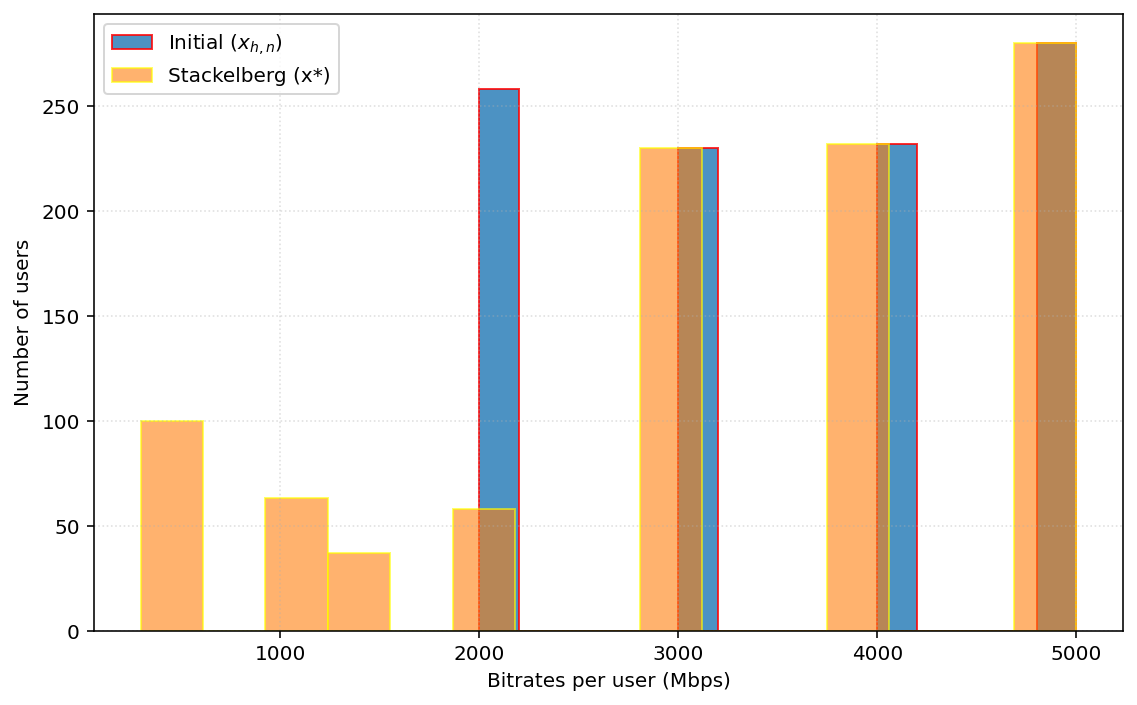}
    \end{array}
    \]
    \caption{Users' bitrates shift from the initial $x_{h,n}$ to $x^\ast$ after the gamified incentivization with $r_n \sim \mathcal{N}(\mu, \sigma^2)$ with budget $B = 1$, for $H = 1$ (left) and $H = 1000$ (right).}
    \label{fig:Stack eq B 1}
\end{figure}

\begin{figure}[t]
    \centering
    \[
    \begin{array}{cc}
    \includegraphics[scale=0.35]{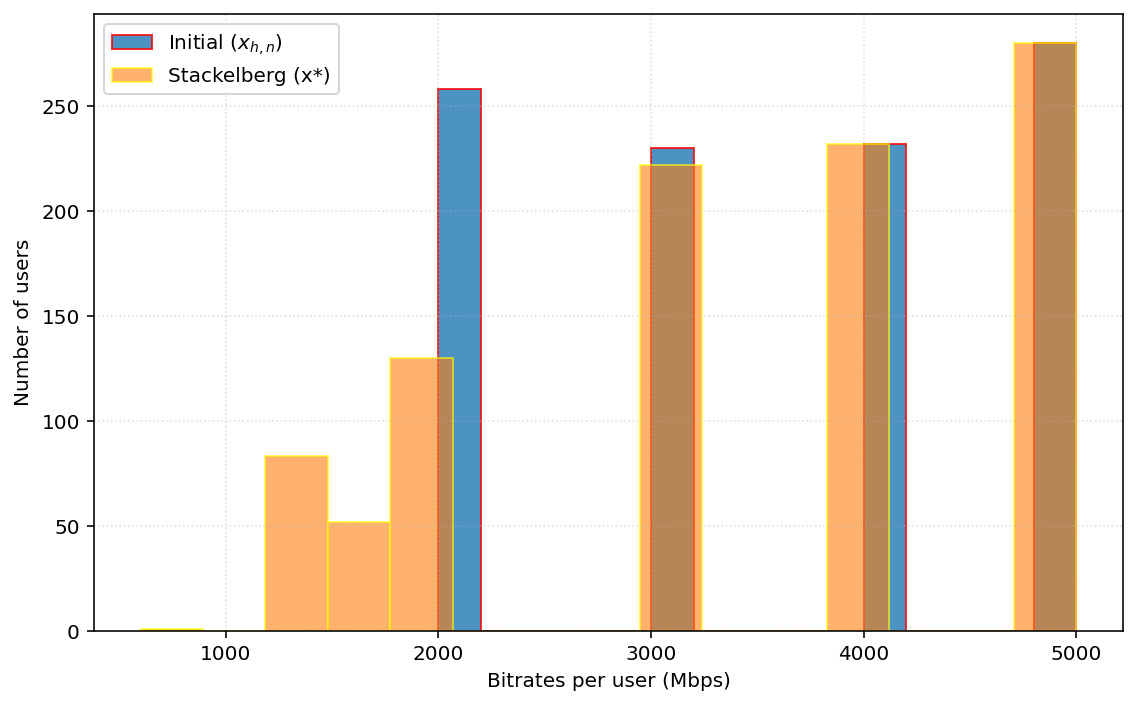} & \includegraphics[scale=0.35]{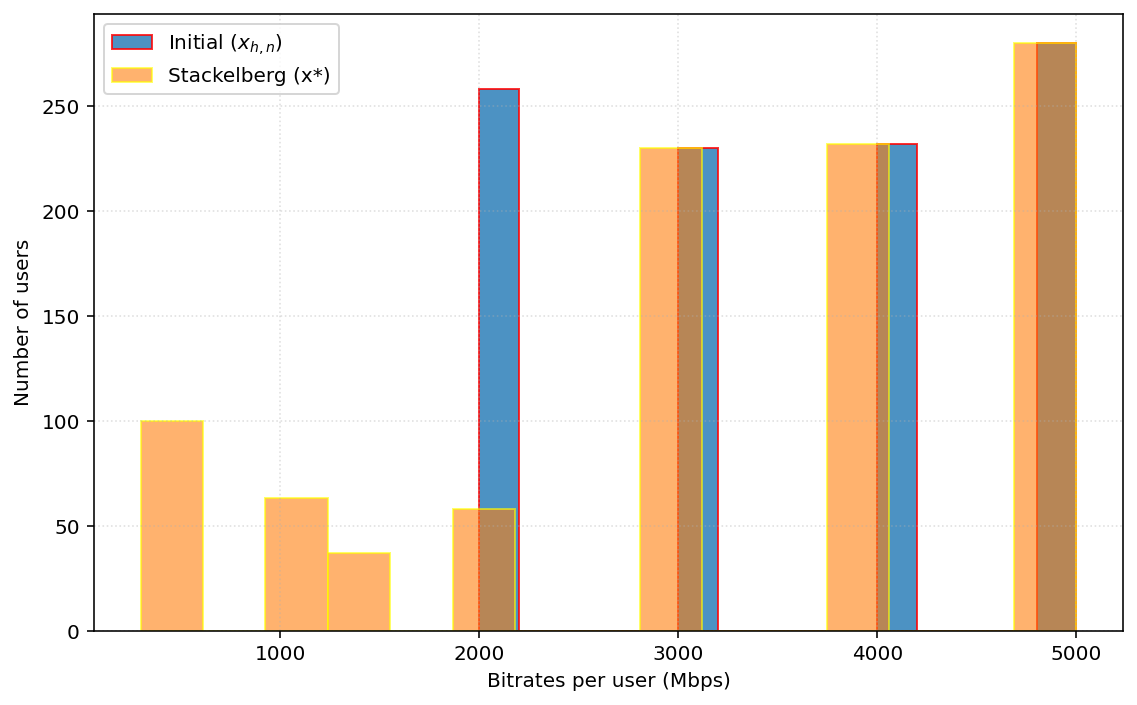}
    \end{array}
    \]
    \caption{Users' bitrates shift from the initial $x_{h,n}$ to $x^\ast$ after the gamified incentivization with $r_n \sim \mathcal{N}(\mu, \sigma^2)$ with budget $B = 100$, for $H = 1$ (left) and $H = 1000$.}
    \label{fig:Stack eq B 100}
\end{figure}

\begin{figure}[t]
    \centering
    \[
    \begin{array}{cc}
    \includegraphics[scale=0.35]{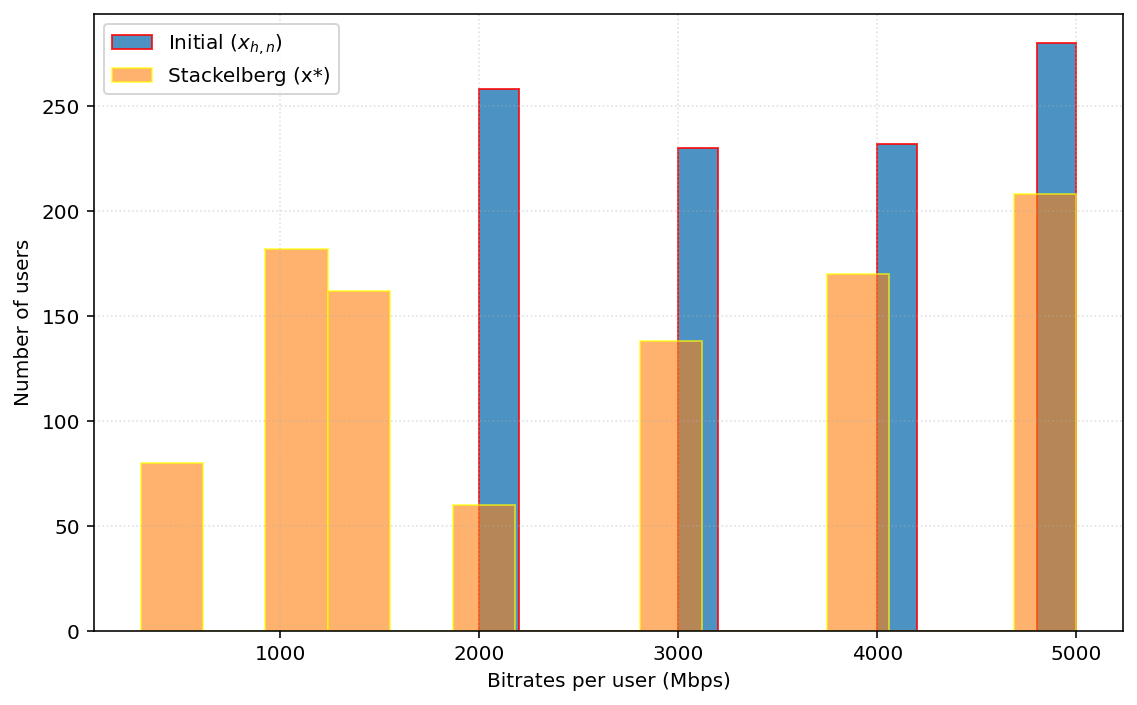} & \includegraphics[scale=0.35]{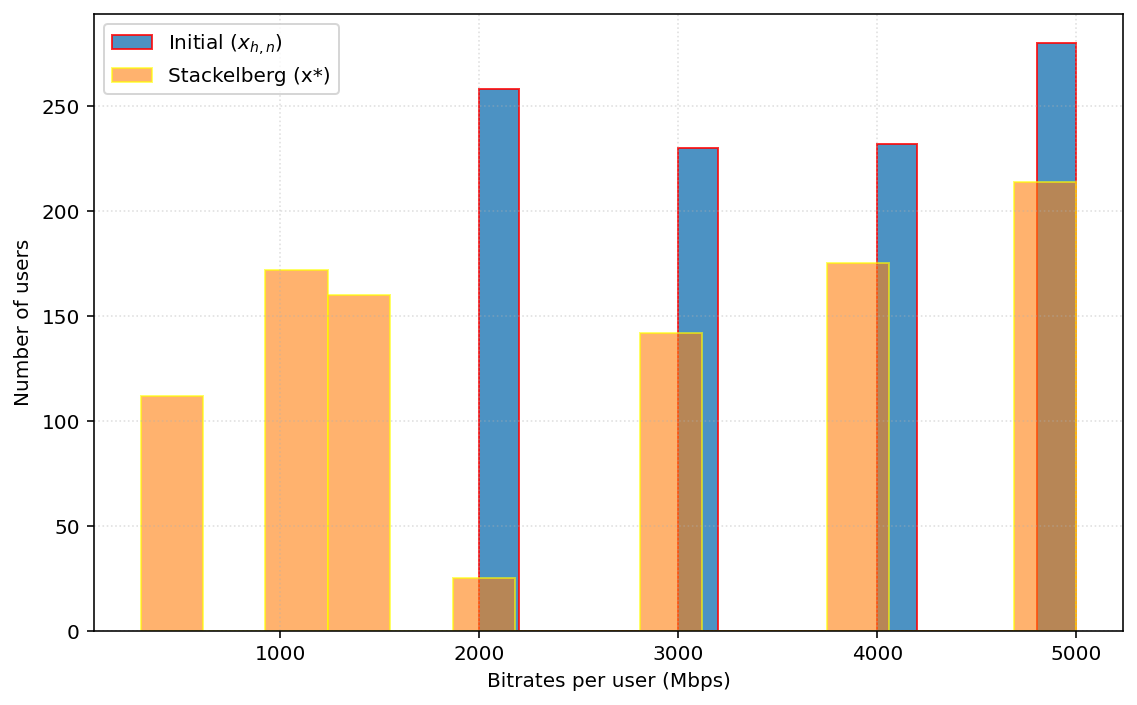}
    \end{array}
    \]
    \caption{Users' bitrates shift from the initial $x_{h,n}$ to $x^\ast$ after the gamified incentivization with $r_n \sim \mathcal{N}(\mu, \sigma^2)$ with budget $B = 1000$, for $H = 1$ (left) and $H = 1000$ (right).}
    \label{fig:Stack eq B 1000}
\end{figure}

Contrarily, in Figure \ref{fig:Percentage Traffic Reduction large rn H 1000}, the range of $B$ is insufficient to address all the offering incentives, since they can become $\approx 10564$ MU. So the energy savings are halted by the limitations in the budget. For this case, experimentally, the system reaches energy reduction, $\approx 73\%$, for $B \approx 12000$, covering almost all the possible energy savings. Interestingly, by the previous analysis, we observe that the available budget needs to be higher than the maximum $\mathbb{E}[r_n]$ MU, which is due to the probabilistic nature of the acceptance. The effect of $B$ is further investigated in Figures \ref{fig:Stack eq B 1}, \ref{fig:Stack eq B 100}, and \ref{fig:Stack eq B 1000}, where we explore the strategic interaction between the provider and the users.
Now, given the information $\mu$ and $\sigma$ for the population, we twist the experiment, and the provider draws the offered incentives from a lognormal distribution $\mathcal{LN}(\nu, \rho^2)$
, with $\nu = \ln(\mu) - \frac{1}{2}\sigma^2$ and $\rho = \sqrt{\ln\left(1 + \frac{\sigma^2}{\mu^2}\right)}$, meaning that it targets the users with higher energy consumption.
\begin{figure}[t]
    \centering
    \includegraphics[scale=0.45]{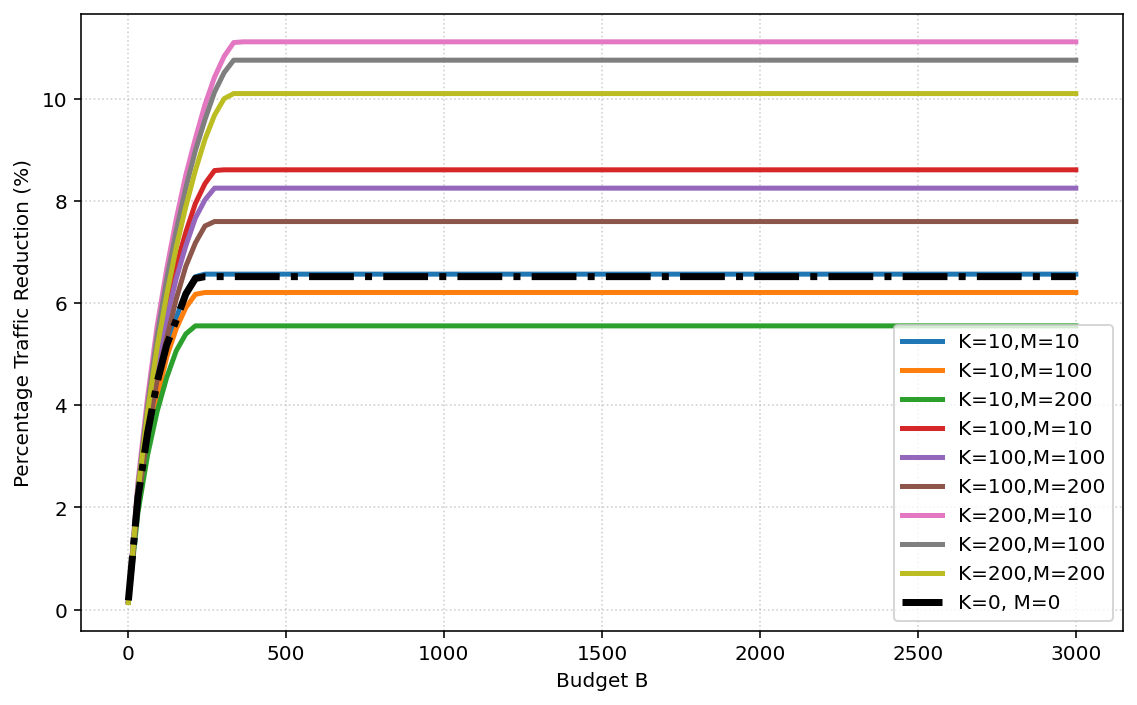}
    \caption{Percentage of traffic reduction with $r_n \sim \mathcal{LN}(\nu, \rho^2)$, when $\mu = 1$ and $\sigma = 0.5$, and $H = 1000$.}
    \label{fig:Percentage Traffic Reduction Log small rn H 1000}
\end{figure}

\begin{figure}[t]
    \centering
    \includegraphics[scale=0.45]{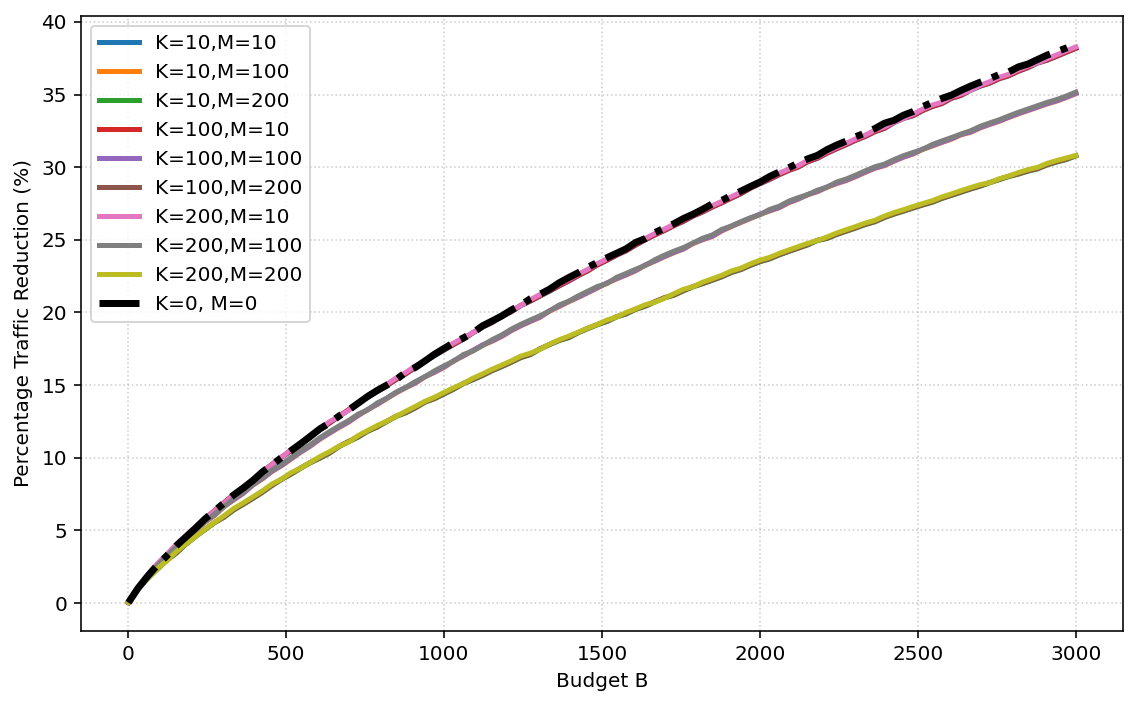}
    \caption{Percentage of traffic reduction for $r_n \sim \mathcal{LN}(\nu, \rho^2)$, when $\mu = 3$ and $\sigma = 2$, and $H = 1000$.}
    \label{fig:Percentage Traffic Reduction Log medium rn H 1000}
\end{figure}

\begin{figure}[t]
    \centering
    \includegraphics[scale=0.45]{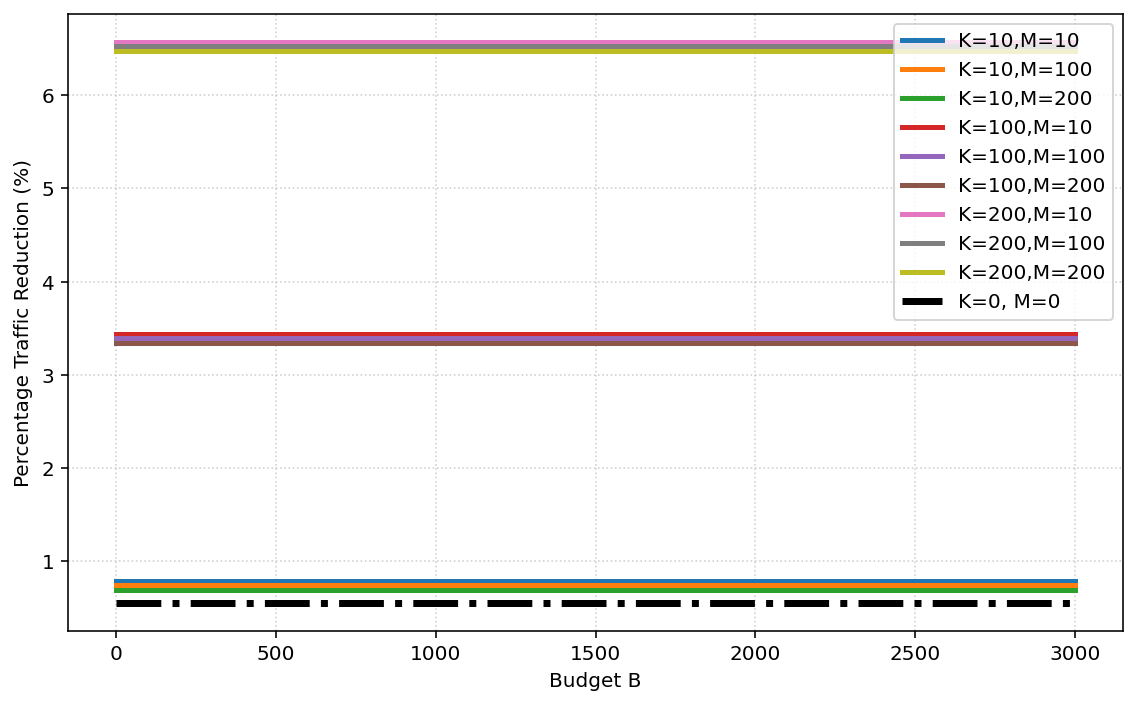}
    \caption{Percentage of traffic reduction for $r_n \sim \mathcal{LN}(\nu, \rho^2)$, when $\mu = 10$ and $\sigma = 4$, and $H = 1000$.}
    \label{fig:Percentage Traffic Reduction Log large rn H 1000}
\end{figure}
In Figure \ref{fig:Percentage Traffic Reduction Log small rn H 1000}, the provider offers total incentives $\approx 980$ MU and misses $85\%$ of the $r_{\min, n}$. As the $r_n$s are small, they are close to $r_n \sim \mathcal{N}(1, 0.25)$, so the energy reduction has the same behavior. In Figure \ref{fig:Percentage Traffic Reduction Log medium rn H 1000}, the provider offers a total incentives drop to $\approx 490$ MU and misses $96\%$ of the $r_{\min, n}$. Therefore, the system's behavior is deteriorated compared to Figure \ref{fig:Percentage Traffic Reduction medium rn H 1000}, where the provider uses population information to sample from $\mathcal{N}(\mu, \sigma^2)$. The situation becomes even worse when $\mu = 10$ and $\sigma = 4$. Consequently, in case the provider has limited information regarding the population, it is most beneficial to pick an unbiased approach, rather than targeting a specific and unknown group of the population.

\begin{table}[t]
    \centering
    \caption{Percentage of traffic reduction for $H = 1000$ and $B=1000$.}
    \label{tab:traffic reduction H = 1000}
    \begin{tabular}{|c|c|c|c||c|c|}
    \hline
    \multicolumn{2}{|c|}{} & \multicolumn{2}{|c||}{$\mathcal{N}(1,0.25)$} & \multicolumn{2}{|c|}{$\mathcal{N}(3,4)$} \\
    \hline
      K & M  & vs $\max \Delta E$ & vs $E(x_{h,n})$ & vs $\max \Delta E$ & vs $E(x_{h,n})$  \\
        \hline
    $0$ &  $0$  & 4.58 & 7.21 &  0.98 & 42.1 \\
    \hline
    \hline
     \multirow{ 3}{*}{$10$} & 10  & 4.48   & 7.12 &  1.50   &  41.6 \\
        \cline{2-6}
          & 100 & 4.90 & 6.78 & 6.80 & 36.3 \\
        \cline{2-6}
        & 200 & 5.53 & 6.58 & 12.8 & 30.9\\
         \hline
        \hline
        \multirow{ 3}{*}{$100$} & 10 & 2.48 & 9.32 & 0.89 & 42.2 \\
        \cline{2-6}
          & 100 &  2.90  &  8.89 &  6.20  &  36.9 \\
        \cline{2-6}
          & 200  & 3.53 & 7.88 & 11.5 & 31.5 \\
         \hline
         \hline
        \multirow{ 3}{*}{$\mathbf{200}^\ast$} & $\mathbf{10}^\ast$ & --            & $\mathbf{11.8}^\ast$ & -- & $\mathbf{43.1}^\ast$ \\
        \cline{2-6}
          & 100  & 0.40 & 11.4 & 5.3 & 37.8  \\
        \cline{2-6}
        & 200 & 1.00 & 10.8 & 10.7 & 32.4 \\
         \hline
    \end{tabular}
\end{table}
Table \ref{tab:traffic reduction H = 1000} refines the insights of Figures \ref{fig:Percentage Traffic Reduction small rn}–\ref{fig:Percentage Traffic Reduction large rn H 1000}. In detail, the combination (large-$K$, small-$M$) with strong social reward $H$ and higher-mean/higher-variance incentive draws produce the largest traffic reductions. The results confirm the characteristic budget–response behavior observed earlier: traffic reduction grows monotonically with $B$. 

Overall, Table \ref{tab:traffic reduction H = 1000} and Figures \ref{fig:Percentage Traffic Reduction small rn}–\ref{fig:Percentage Traffic Reduction large rn H 1000}, substantiate that increasing $B$ enhances network-wide efficiency only up to the point where behavioral saturation limits further gains, and that the asymmetric gamification design (large-$K$, small-$M$) remains the most cost-effective strategy for low-monetary-reward environments. Therefore, a practical provider's policy could (i) first apply asymmetric gamification and tune offered-incentive distribution (mean sufficiently large); (ii) increase budget only until it covers the meaningful acceptance mass.

\paragraph*{Stackelberg equilibrium.} In Figures \ref{fig:Stack eq B 1}, \ref{fig:Stack eq B 100} and \ref{fig:Stack eq B 1000}, we present the outcome of the serious game, as defined in equation \eqref{eq:Stack-eq}. In this experiment, the provider decides regarding the offered incentives $r_n \sim \mathcal{N}(\mu, \sigma^2)$ and values for the $K$ and $M$. So, the strategic space for the provider is $K \times M \times \mu \times \sigma$, where $K, M \in \{0, 10, 50, 100, 150, 200\}$, $\mu \in \{1, 2, 3, 4, 6\}$ and $\sigma = \{0.5, 1, 2, 3\}$, and $\{x_{h,n}, x_{l, n}\}$ for the user $n$. Further, we assume $H \in \{1, 1000\}$, and $B \in \{1, 100, 1000\}$. 

For $B=1$, Figure~\ref{fig:Stack eq B 1}, the provider reaches equilibrium at $(r_n^\ast,K^\ast,M^\ast)=(200,10,\mathcal{N}(1,1))$ when $H=1$, resulting in $94$ users ($9.4\%$) switching to the lower bitrate and a system-wide reduction of approximately $10$~kW/h. When $H=1000$, the equilibrium shifts to $(200,10,\mathcal{N}(1,0.5))$, with $200$ users ($20\%$) switching and a total reduction of $\approx 44.4$~kW/h.

For $B=100$, Figure~\ref{fig:Stack eq B 100}, the provider selects $(200,10,\mathcal{N}(1,0.5))$ for both $H=1$ and $H=1000$. Under $H=1$, $135$ users ($10.2\%$) switch, yielding a reduction of $\approx 12.6$~kW/h, while for $H=1000$, $200$ users ($20\%$) switch, reducing consumption by $\approx 44.4$~kW/h. The equilibrium here coincides with the previous case, suggesting that the dominant factor in these two settings is the social reward $H$.

Finally, for $B=1000$, Figure~\ref{fig:Stack eq B 1000}, the equilibria are $(200,10,\mathcal{N}(2,0.5))$ for both $H=1$ and $H=1000$, with $424$ users ($43.4\%$) switching and $\approx 144$~kW/h reduction, and  $444$ users ($44.4\%$) switch and the reduction reaches $\approx 158$~kW/h, respectively.

The gamification and offered incentives push the users with lower/moderate high bitrates to alter their choices, but need careful treatment. In the left panels of Figures \ref{fig:Stack eq B 1}-\ref{fig:Stack eq B 1000}, gamification has little effect, because the rewards $h_n$ are too low. However, since they are closer to the $x_{\ell, n}$ bitrates, the scale of $r_n$, $\delta_n$, and the $\gamma_n$ may encourage greener behavior. In the right panels of Figures \ref{fig:Stack eq B 1}-\ref{fig:Stack eq B 1000}, with a high reward $H = 1000$, gamification motivates more users to alter their choices. In general, users with higher high bitrates are unaffected by low $h_n$, particularly when $r_n$ is small; to influence them, the provider must either increase $r_n$ or the environment must offer larger rewards. This also explains why, when $H$ is very large, the equilibrium can remain unchanged for a wide range of $B$, since social rewards dominate monetary incentives.

Moreover, Figures \ref{fig:Stack eq B 1}-\ref{fig:Stack eq B 1000} illustrate how the Stackelberg equilibrium evolves as the provider’s budget $B$ increases. In particular, the feasible space of the provider’s incentive policy expands, allowing it to configure $(r_n,K, M)$ to align with users’ heterogeneous responsiveness profiles. What emerges is a transition from a \emph{budget-constrained} regime (low $B$)--where only highly or socially motivated users switch--to a \emph{behavior-constrained} regime (high $B$) where many responsive users have already shifted to the greener option. In this progression, budget indirectly alters switching patterns by reshaping the distribution and variance of offered incentives, which modifies the effective slope of the acceptance $p_n(r_n; K, M)$.

\begin{figure}[t]
    \centering
    \includegraphics[scale=.7]{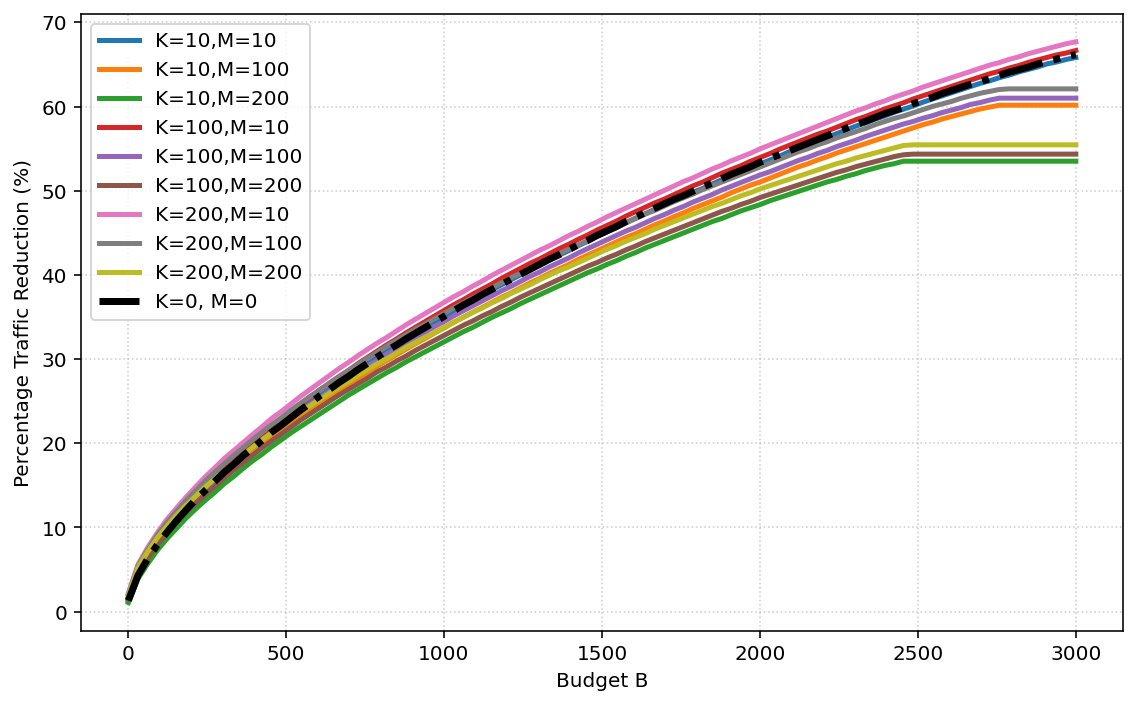}
    \caption{Traffic reduction from Ultra-HD to Full-HD, when $r_n \sim \mathcal{N}(\mu = 3, \sigma = 2)$, and $H = 1000$.}
    \label{fig:Traffic Reduction Ultra to Full}
\end{figure}

\paragraph*{From Ultra-HD to Full-HD.} Now we illustrate the applicability of our framework to a real-world scenario. Assume that each of the $N$ users consumes $20$~Mbps, that is $x_{h,n} = 20000$, and there is a greener option with $5$~Mbps, that is $x_{l,n} = 5000$, for each $n \in N$. This setup can be viewed as a typical video-streaming scenario where each household or user watches $20$~Mbps Ultra-HD (4K) content, and a greener alternative corresponds to switching to $5$~Mbps Full HD (1080p). The maximum available traffic reduction is about $\approx 75\%$ of the situation where all the users use the $x_{h,n}$ bitrates. Utilizing the gamified incentives model we achieve a traffic reduction up to $\approx 67.2\%$, see Figure \ref{fig:Traffic Reduction Ultra to Full}. This indicates that the proposed incentive mechanism can exploit most of the theoretical reduction potential even in realistic, high-demand conditions. In practice, it suggests that by combining monetary and social rewards, a provider could motivate users to downgrade video quality from Ultra-HD to Full-HD while retaining nearly 90\% of the maximum achievable traffic reduction. \\

Importantly, gamification (large $K$, small $M$, high $H$) amplifies the budget’s impact by substituting social reward for monetary cost, steepening users’ response curves, and accelerating the system’s convergence toward equilibrium. In general, increasing $B$ strengthens the provider’s control only up to the point where user heterogeneity and saturation of acceptance probabilities, rather than financial limitations, become the dominant constraints on energy reduction. Moreover, across all experiments, the provider can improve the outcome by adjusting $\mu$ and $\sigma$, meaning that the situation evolves in multiple turns.

\noindent Concluding, the proposed framework empowers the service provider to influence user behavior—such as choosing lower video resolutions or potentially shifting usage away from peak hours—resulting in traffic and energy reductions that benefit the entire system. Traffic reduction eases congestion and improves scalability for Content Delivery Networks (CDNs) and Network Service Providers (NSPs), while energy savings lower operational costs and emissions for service hosts and energy suppliers. By dynamically adjusting incentives and game parameters, the provider aligns user actions with system-wide efficiency goals, fostering a collaborative environment of improved performance and greener operations.

\section{Conclusions}\label{sec:conclusions}

This paper presents an energy-aware service management framework that blends personalized incentives with social engagement, using greenness-aware QoE modeling and game-like mechanisms (e.g., top-K and bottom-M rankings). The framework enables service providers to shift from passive incentives to strategic, application-level control. Using a Stackelberg game model, we optimize incentives and game parameters within budget limits, harnessing social influence to drive behavior with less dependence on monetary rewards. The proposed approach achieves up to 67.2\% traffic reduction, over a factor-3 decrease. Gamification amplifies the budget’s impact by substituting social reward for monetary cost. The proposed model lays a foundation for future research on fairness, long-term behavioral adaptation, and policy-aligned incentive design in ICT services.

\newpage

\section*{Acknowledgment}

This work has been partly developed in the scope of the project EXIGENCE, which has received funding from the Smart Networks and Services Joint Undertaking (SNS JU) under the European Union (EU) Horizon Europe research and innovation programme under Grant Agreement No 101139120. Views and opinions expressed are however those of the author(s) only and do not necessarily reflect those of the EU or SNS JU.


\bibliographystyle{plain} 
\bibliography{refs} 

\end{document}